\documentclass[twocolumn,prl]{revtex4}
\usepackage{graphicx}
\usepackage{amsmath,amsthm,bbm,times,mathtools}
\usepackage{amssymb}
\usepackage{bm}
\usepackage{bbold}
\usepackage{color}

\newtheorem{theorem}{Theorem}
\newcommand{\tr}{\mathrm{Tr\,}}

\newcommand{\e}{\textnormal{e}}
\newcommand{\dd}{\textnormal{d}}
\def\E{\mathbbm{E}}
\def\ceq{\coloneqq}

\date{12 November 2021, appeared as editors' suggestion in Phys. Rev. Lett. \textbf{127}, 207204, 6 pp. (2021)}

\begin{document}

\title{Existence of replica-symmetry breaking in quantum glasses}

\author{Hajo Leschke}
\affiliation{Institut f\"ur Theoretische Physik, Universit\"at Erlangen--N\"urnberg, 91058 Erlangen, Germany}

\author{Chokri Manai} 
\affiliation{Department of Mathematics and Munich Center for Quantum Science and Technology, TU M\"{u}nchen, 85747 Garching, Germany}

\author{Rainer Ruder}
\affiliation{Institut f\"ur Theoretische Physik, Universit{\"a}t Erlangen--N\"urnberg, 91058 Erlangen, Germany}

\author{Simone Warzel}
\affiliation{Departments of Mathematics and Physics, Munich Center for Quantum Science and Technology, TU M\"{u}nchen, 85747 Garching, Germany}

\begin{abstract}
	By controlling quantum fluctuations via the Falk--Bruch inequality we give the first rigorous argument for the existence of a spin-glass phase in the quantum Sherrington--Kirkpatrick model with a ``transverse'' magnetic field if the temperature and the field are sufficiently low. The argument also applies to the generalization of the model with multi-spin interactions, sometimes dubbed as the transverse $p$-spin model. 
\end{abstract}

\maketitle

\paragraph{Introduction.} 
Spin glasses constitute a particularly multi-facetted topic in the statistical mechanics of disordered systems.
Classical spin-glass models, such as the mean-field one by Sherrington and Kirkpatrick (SK) \cite{SK75}, were originally introduced to understand the unusual magnetic properties observed in some metal alloys with irregularly competing ferro- and anti-ferromagnetic interactions.
Beyond their ongoing significance in condensed-matter physics \cite{FH91}, such models with their built-in frustration have evolved meanwhile into paradigms in optimization, information processing, and the theory of neural networks~\cite{N01,MM09}.
Their rich low-energy structure and complexity continues to generate deep scientific discoveries. For example, among the recent excitements in computation is a conditional proof (based on the widely believed assumption of $\infty $-replica-symmetry breaking) of the existence of a polynomial-time classical algorithm for finding an approximate bit-string whose energy is with high probability $ \varepsilon $-close to the lowest SK energy \cite{M21}.
Such an algorithm is not believed to exist for a search of the ground-state energy in $p$-spin generalizations of the SK model.
Quantum mechanics promises to offer help in the form of quantum adiabatic annealing or quantum approximate optimization algorithms  \cite{B+13,DA+15,K16,AL18,BL18,FGGZ19,C+21}.
In this context, but also purely motivated by the fact that spin glasses are prototypes for the emergence of non-ergodic behavior in disordered quantum systems~\cite{RCC89,LPS14,BL+17,MC19,S+20}, it is important to study quantum versions of classical spin-glass models. This can be done by taking the quantum nature of spins seriously and by adding a ``transverse" magnetic field to the classical energy landscape, which induces quantum effects.
Most prominent is the  \emph{quantum Sherrington--Kirkpatrick model} (QSKM) with $N\geq 2$ three-component vector spins of main quantum number $1/2$ (or qubits). Their $z$-components interact with each other in a random fashion, while their $x$-components interact individually with a constant magnetic field of strength $b\geq0$ externally applied along the positive $x$-direction. Up to a factor $1/2$, the $j$-th spin operator may be represented by the triple 
\[
	S_j^x = \left(\begin{matrix} 0 & 1 \\ 1 & 0 \end{matrix} \right)\,, \quad S_j^y =  \left(\begin{matrix} 0&- i \\ i & 0 \end{matrix}\right)\,, \quad
	S_j^z = \left(\begin{matrix} 1 & 0 \\ 0 & -1 \end{matrix}\right)
\]
of Pauli matrices and is meant to act on the $j$-th factor of the tensor-product Hilbert space $\mathcal{H}  _N \ceq \otimes_{j=1}^N \mathbbm{C}^2 $ and as the identity on the other factors. The  Hamiltonian (or energy operator) of the QSKM is then defined on $\mathcal{H}_N$ by the sum
\begin{equation}\label{H}
	H_N \ceq J \ U_{N}- b \sum_{j=1}^N S^x_j\quad,\quad J>0\,,\quad b\geq0\,,
\end{equation}
with its (dimensionless) classical zero-field SK part 
\begin{equation}\label{H0}
	U_{N}\ceq -\frac{1}{\sqrt{N}}\sum_{1\leq j<k\leq N}  g_{jk} S^z_j S^z_k\,.
\end{equation}
Here the spin coupling is (only) pairwise and given by independent, identically distributed Gaussian random variables $ (g_{jk})$ with mean $\E[g_{12}]=0$ and variance $\E\big[ g_{12}^2\big]=1$, modeling frozen-in spatial disorder of the glass of strength $J>0$.

As usual, the thermal average for reciprocal temperature $\beta\in{]0,\infty[}$ is given by the canonical \emph{Gibbs expectation} $\langle\,\cdot\,\rangle\ceq \tr \e^{-\beta H_N} (\,\cdot\,) / Z_N$ with the partition function $Z_{N}\ceq \tr\e^{-\beta H_N}$ as the normalization factor. For $b=0$ there is no a-priori ``globally" preferred spin orientation and no conventional magnetic order arises. Yet, one expects spin-glass order even for $b\geq0$ in the sense that $\E[q_N]=\E\big[\langle S_1^{z}S_2^{z}\rangle^2\big]>0$ in the limit of a ``macroscopically" large number of spins ($N\to\infty$), provided that the temperature and the field are sufficiently low. Here we are using the model's spin-index symmetry under the (probabilistic) \emph{disorder expectation} $\E[\,\cdot\,]$ and the ${[0,1]}$-valued random variable
\begin{equation}\label{OP}
	q_N \ceq \frac{2}{N(N-1)}\sum_{1\leq j<k\leq N} \langle S^z_j S^z_k \rangle^2
\end{equation}
as the corresponding order parameter. It may be rewritten as
\[
	q_N=\frac{N}{N-1}\langle R_N^2 \rangle^{\otimes}-\frac{1}{N-1}
\]
in terms of $R_N\ceq N^{-1}\sum_{j=1}^N S_j^z\otimes S_j^z$, the \emph{replica-over\-lap operator}  for the ``duplicated model" with Hilbert space $\mathcal{H}_N\otimes\mathcal{H}_N$, Hamiltonian $H_N\otimes\mathbbm{1}+\mathbbm{1}\otimes H_N$, and associated Gibbs expectation $\langle\,\cdot\,\rangle^{\otimes}$. Strict positivity of $\E[q_N]=\E\big[\langle S_1^{z}S_2^{z}\rangle^2\big]$ is therefore equivalent to replica-symmetry breaking (as $N\to\infty$).\\

\paragraph{Main result.} The main result of this Letter is a proof of this replica-symmetry breaking at small enough temperature and field strength. This is facilitated by extending a key observation of Bray and Moore \cite{BM80}, generalized to certain non-Gaussian probability distributions of $g_{12}$ by Aizenman, Lebowitz, and Ruelle~\cite{ALR87}, to the present quantum case $b>0$: the mean order parameter $\E[q_N]$ is related to the mean $\E\big[\langle U_{N}\rangle\big]$ of the zero-field part of the Hamiltonian. Specifically, by the spin-index symmetry and a standard Gaussian integration by parts it is straightforward to obtain
\begin{align}\label{eq:ALR}
	&-\frac{2}{N-1}\E\big[\langle U_{N}\rangle\big] = \sqrt{N}\,\E\big[g_{12}\,\langle S^z_1 S^z_2\rangle\big]\notag \\
	& = \sqrt{N}\,\E\big[\partial\langle S^z_1 S^z_2\rangle\big/\partial g_{12}\big] \notag \\
	& = \beta J\,\E\big[\langle S^z_1 S^z_2|S^z_1 S^z_2\rangle - \langle S^z_1 S^z_2\rangle^2\big]\notag \\ 
	& = \beta J\,\E\big[\langle A|A \rangle -\langle A \rangle^2\big]
\end{align}
in terms of the observable $A\ceq S^z_1 S^z_2$ and its Duhamel--Kubo--Bogolyubov scalar product \cite{DLS78,KTH98} with itself:
\[
	\langle A|A\rangle\ceq\int_0^1\!\!\dd t\,\big\langle e^{t\beta H_N} A^{*} e^{-t\beta H_N} A \big\rangle\,.
\]
It satisfies the well-known a-priori estimates $0\leq\langle A \rangle^2\leq\langle A|A \rangle\leq\langle A^2\rangle=1$, where the inequalities hold for general (self-adjoint) $A=A^{*}$ and the equality is due to $A^2=\mathbbm{1}$ for the present $A$. In the classical commutative case, $b=0$, the third inequality is also an equality and \eqref{eq:ALR} turns into (4.3) of \cite{BM80} and (4.1) of \cite{ALR87} (for Gaussian disorder).

For general $b\geq 0$ we need a lower bound on $\langle A|A \rangle$ better than $\langle A \rangle^2$ in order to obtain a non-trivial lower bound on $\E\big[\langle A \rangle^2\big]$ from \eqref{eq:ALR}. As our second main ingredient for the proof, we control the quantum fluctuations by the Falk--Bruch inequality \cite{FB69} (see also \cite{R77,DLS78}):
\begin{equation}\label{FB}
	\langle A|A\rangle \geq \langle A^2\rangle \, \Phi\Big(\frac{1}{4 \langle A^2\rangle}\big\langle \big[A,[\beta H_N,A]\big]\big\rangle \Big)\,.
\end{equation}
The function $\Phi:{[0,\infty[} \to {]0,1]}$ from the positive half-line to the left-open unit interval is defined implicitly by the relation $\Phi\big(r\tanh(r)\big)\ceq r^{-1}\tanh(r)$. It is monotone-decreasing and convex with $\Phi(0)=1$. Moreover, it can be estimated from below according to $\Phi(t)\geq t^{-1}(1-e^{-t})\geq \max\{0,1-t/2\}$, see \cite{DLS78}. We also note that the Gibbs expectation of the double commutator 
in the argument of $\Phi$
in \eqref{FB} equals the scalar product $\big\langle[\beta H_N,A]\big| [\beta H_N,A]\big\rangle$ and is hence positive for a general self-adjoint $A$. Since in the present case $A=S^z_1 S^z_2$ commutes with $U_{N}$, the double commutator is independent of $J$ and simply given by
\begin{equation}\label{eq:commutator}
	\big[A,[\beta H_N,A]\big]= 4\beta b (S_1^x + S_2^x)\,.
\end{equation}
Combining \eqref{eq:ALR}, \eqref{FB}, \eqref{eq:commutator}, and using Jensen's inequality for the convex $\Phi$ together with spin-index symmetry yields the basis for our main result:
\begin{theorem}\label{thm:QSK}
The mean of the spin-glass order parameter \eqref{OP} has a lower bound according to
\begin{equation}\label{eq:main}
	\E[q_N] \geq \Phi\Big(2\beta b\,\E\big[\langle S_1^{x} \rangle\big]\Big) + \frac{2}{\beta J} \frac{1}{N-1} \E\big[\big\langle U_{N}\big\rangle \big]\,.
\end{equation}
It is valid for any $\beta>0$, $J>0$, $b\geq0$, and all $N\geq2$.
\end{theorem}

For more explicit bounds we further estimate the right-hand side (RHS)
 of~\eqref{eq:main} starting with its  first term.  Adding to the Hamiltonian~\eqref{H} the term $(b-b_1)S_1^x$ with $b_1\geq0$ and estimating the associated ``local" susceptibility results in the differential inequality  for the transverse magnetization
\[
	\frac{\partial}{\partial b_1}\langle S_1^{x}\rangle_{b_1}\!=\beta\big(\langle S_1^{x}|S_1^{x}\rangle_{b_1} - \langle S_1^{x}\rangle_{b_1}^2\big) \leq \beta\big(1-\langle S_1^{x}\rangle_{b_1}^2\big)\, .
\] 
Integrating by separation of variables and observing $\langle S_1^{x} \rangle_{0} = 0 $, we hence obtain
  $\langle S_1^{x}\rangle\leq\tanh(\beta b)$, which by the monotonicity of $\Phi$ results in the estimates
\begin{equation}\label{eq:th} 
 \Phi\Big(2\beta b\,\E\big[\langle S_1^{x} \rangle\big]\Big) \geq \Phi\big(2\beta b\tanh(\beta b)\big) \geq \Phi\big(2\beta b)\, . 
\end{equation}
A simple bound on the second term in~\eqref{eq:main}  results from  the (non-random) ground-state energy $-\kappa J <0$  of $JU_{N}/N$ as $N\to\infty$  with the constant $\kappa\approx 0.763$ according to \cite{P80b,CR02}. Combined with~\eqref{eq:th} this leads to the more explicit lower bound
\begin{equation}\label{EX}
	\overline{q}(\beta J, \beta b ) \ceq \liminf_{N\to\infty} \E[q_N]  \geq \Phi\big(2\beta b\tanh(\beta b)\big) - \frac{2\kappa}{\beta J}
\end{equation}
on the lower limit of the sequence $\big(\E[q_{N}]\big)_{N\geq2}$ in the unit interval ${[0,1]}$.
For $b=0$ the RHS of \eqref{EX} is strictly positive for temperatures below $ J /(2 \kappa ) \approx 0.655\, J$. This (not maximum) temperature regime for the existence of a spin-glass phase agrees with the one found in (4.14) of \cite{ALR87}. In this regime the spin-glass phase is seen to survive when turning on the transverse magnetic field, provided that $b/J>0$ is so small that the RHS of \eqref{EX} remains strictly positive. This condition is implied by the slightly stronger but simpler one $ 1-\e^{-2\beta b} > 4\kappa b/J $, yielding in the zero-temperature limit the same maximum field strength $J/(4\kappa) \approx 0.328\,J$ as from \eqref{EX}.

To establish the persistence of spin-glass order for sufficiently small $ b/J $ also for temperatures up to the zero-field critical (freezing) temperature $J$, we start from the observation that \eqref{FB} and hence \eqref{eq:main} are equalities for $b=0$ and remain rather sharp for small $\beta b>0$. Consequently, \eqref{eq:main} should cover the whole regime $\beta b\ll 1\leq\beta J$. 
To confirm this, we estimate the mean 
$
\overline{u}(\beta J, \beta b) \ceq \liminf_ {N\to \infty}\E\big[ \big\langle U_{N}\big\rangle \big]/N
$ 
of the  zero-field SK part ~\eqref{H0} by the Fisher-type \cite{F65} inequality
\begin{align}\label{eq:Fisher}
	& \overline{u}(\beta J, \beta b)+a^{-1} \ln\big(\cosh(\beta b)\big) \geq  \overline{u}(\beta J+a, 0) \notag \\
	&= \big(\overline {q}(\beta J+a,0)-1\big)(\beta J + a)/2
 \end{align}
with an arbitrary $a>0$. 
It results from the convexity of $ \ln\big(Z_N(\beta J, \beta b)\big) $ in $ \beta J $ together with  the Peierls--Bogolyubov and Golden--Thompson 
bounds
 $ Z_N(\beta J, 0) \leq Z_N(\beta J, \beta b) \leq Z_N(\beta J, 0)\big(\cosh(\beta b)\big)^N $ on the partition function. 
The equality in \eqref{eq:Fisher} is due to \eqref{eq:main} for $b=0$. 
Using~\eqref{eq:th} and \eqref{eq:Fisher} with $ a=a_b \ceq \sqrt{2 \ln\big(\cosh(\beta b)\big) }$\, in~\eqref{eq:main} for $N\to \infty$ leads to 
\begin{align}\label{eq:cont}
\overline{q}(\beta J, \beta b ) \geq & \Big( 1+ \frac{a_b}{\beta J} \Big) \overline{q}(\beta J + a_b ,0)\notag \\&- \Big( 1 - \Phi\big(2\beta b\tanh(\beta b)\big) + \frac{2a_b}{ \beta J}\Big) \notag \\
\geq  & \, \overline{q}(\beta J + a_b ,0) - 3\beta b \,.
\end{align}
The simplifying second inequality follows by observing $a_b\in{[0,\beta b]}$, estimating $\Phi(t)$ as above, and assuming $ \beta J\geq1 $. Finally,  we fix an arbitrary $\beta J>1$ which is equivalent to $ \overline{q}(\beta J ,0)>0$ characterizing the spin-glass phase for $b=0$, see \footnote{This well-known equivalence follows easily from the inequality $\int_0^{\beta J}\!\!\dd t\,t\,\overline{q}(t,0) \geq 2k(\beta^{2}J^{2}/4)$ for any $\beta J>0$. It is due to \eqref{eq:main} and Guerra's observation \cite{G01} that the (replica-symmetric) SK approximation \cite{SK75} provides a lower bound on $-\E\big[\ln\big(Z_{N}(\beta J,0)\big)\big]/(N\beta)$ for any $N\geq2$. The differentiable function $\lambda\mapsto k(\lambda)$ is zero for $\lambda\leq1/4$ and strictly positive and increasing for $\lambda>1/4$, see \cite{SK75,LRRS21}.}. The continuity of $\overline{u}(\beta J+a,0)$ in $a$ (by \cite{T06c,P08}) and hence of $\overline{q}(\beta J+a, 0)$ yields the continuity of the RHS of \eqref{eq:cont} in $\beta b$. Its strict positivity for $b=0$ therefore extends to sufficiently small $\beta b \in {]0,1/3[}$. In other words, the well-known spin-glass phase without a field persists with a low enough transverse field at \emph{any} temperature below $J$.

\begin{figure}[h]
\!\!\!\!\!\!\includegraphics{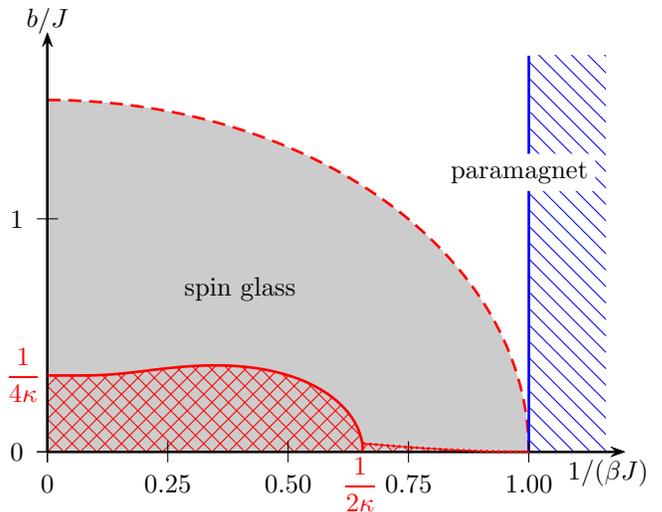}
\caption{In the temperature-field plane the (red) cross-shaded regime indicates where we prove the existence of spin-glass order in the QSKM by \eqref{EX} and,  respectively, by~\eqref{eq:cont} combined with~\cite{S85} (see text). 
The (red) dashed line is a cartoon of the critical line between the spin-glass and the paramagnetic phases as obtained by approximate arguments and/or numerical methods \cite{FS86,US87,YI87,GL90,Y17,MRC18}. The (blue) line-shaded regime for $\beta J<1$ indicates where the spin-glass order parameter is rigorously known to vanish \cite{LRRS21}.\label{regionplot}}
\end{figure}

\paragraph{Discussion.}
Over the years various approximate and/or numerical studies like \cite{FS86,US87,YI87,GL90,Y17,MRC18} have suggested for the QSKM a temperature-field phase diagram with a critical line between the spin-glass and the paramagnetic phases as sketched in Fig.~\ref{regionplot}, see also \cite{S+13}. In particular, these studies have predicted a quantum phase transition at zero temperature and $b/J \approx 1.51$ or $1.6$.
The (red) cross-shaded regime in Fig.~\ref{regionplot} illustrates where we prove the existence of spin-glass order by the lower bounds \eqref{EX} and \eqref{eq:cont}. Here, the tiny regime above the temperature $J/(2\kappa)$ is produced by inserting the asymptotic expansion of $\overline{q}(\beta J ,0)$ close to $\beta J = 1$ from~\cite{S85} into the RHS of~\eqref{eq:cont}. Apart from that we have no prediction for the location of the true critical line. In particular, our zero-temperature ``critical" field $J/(4\kappa)$ is very likely too small, as is the whole cross-shaded regime. The precise location and nature of the true quantum critical point remains an important problem, in particular in the context of adiabatic algorithms. Nevertheless, our rigorous result supports the conjecture that the ground state typically has localization properties with respect to the eigenbasis of $U_N$. It does not rule out, though, a weak form of restoration of ergodicity through quantum tunneling for those parameters put forward in~\cite{RCC89,BU90,MRC18,MC19}. To clarify this question it is necessary to consider the probabilistic distribution function of the order parameter and not just its mean, because the sequence $(q_N)_{N\geq 2}$ is not expected to be self-averaging in the spin-glass phase.

In this context, we recall that for $b=0$ the mean free energy, $-\lim_{N\to \infty}\E\big[\ln\big(Z_N(\beta J,0)\big)\big]/(N\beta)$, and hence also the RHS of~\eqref{eq:Fisher}, is exactly determined by Parisi's (zero-field) distribution function on ${[0,1]}$ which with increasing $\beta J>1$  exhibits $\infty $-replica-symmetry breaking  \cite{P80a,P80b,CR02,T06, T06c, P08}. 
In contrast, no closed-form expressions are available for $b>0$. Recently the QSKM free energy, which previously has been proved to exist and to be independent of the specific probability distribution of the coupling coefficient $g_{12}$ as long as $ \E[g_{12}] = 0 $ and $ \E\big[|g_{12}|^{3}\big]<\infty$ (see \cite{C07}), was shown to be given by a variational formula in terms of a Parisi-like functional for an infinite-component vector-spin model \cite{AB20}. However, no conclusion could so far be drawn about emerging phases form this formula. In contrast, for the simpler case $\beta J<1$ it is known \cite{LRRS21} that the free energy coincides with its annealed version and that there is no spin-glass phase for any $b\geq 0 $, see the (blue) line-shaded regime in Fig.~\ref{regionplot}. The combination of this result with the present one rigorously proves the existence of a phase transition in the QSKM related to replica-symmetry breaking. But Fig.~\ref{regionplot} clearly calls for further rigorous work on this model.

For a family of quantum hierarchical models dubbed as QGREM, which for $b=0$ were originally introduced by Derrida \cite{D85} as approximations to the more difficult SK model, explicit formulas for the free energy are available \cite{MW21} also for $b>0$. Unlike for their classical counterparts, the phase diagrams of these QGREMs seem to capture the QSKM only on a qualitative level though, since their critical lines reach up to $\beta J=0$ separating a quantum paramagnetic phase from a classical one at high temperatures.\\

\paragraph{Extensions.} The above simple strategy for proving replica-symmetry breaking has straightforward extensions. From our proof it is evident that Theorem~\ref{thm:QSK} remains true as it stands if one adds to \eqref{H} any term commuting with $U_{N}$ that is possibly random but independent of $U_{N}$ such as, for example, a Zeeman term corresponding to a magnetic field in $z$-direction. Adapting the more-involved argument of \cite{ALR87}, our bounds can also be extended from Gaussian to more general symmetric distributions of the coupling coefficients.

This strategy can also be applied to quantum spin-glass models with multi-spin interactions, for example to the ``transverse $p$-spin model". This model generalizes the zero-field SK part \eqref{H0} of \eqref{H} for each natural $p\geq 2$ to
\[
	U_N = -\sqrt{\frac{p!}{2N^{p-1}}}\mkern-15mu\sum_{1\leq j_1<\dots< j_p\leq N}\mkern-15mu g_{j_1j_2\dots j_p} S^z_{j_1}S^z_{j_2} \cdots S^z_{j_p}\,,
\]
where $(g_{j_1j_2\dots j_p})$ are independent and identically distributed standard Gaussian random variables. For $p>2$ this classical zero-field Hamiltonian exhibits at its freezing temperature finite and not $\infty $-replica-symmetry breaking \cite{G85}. Proceeding for the quantum model as in \eqref{eq:ALR} and introducing $\alpha_p(N) \ceq N!\big/\big[(N-p)!\,N^p\big]$, which tends to one as $N\to\infty$, the mean zero-field energy 
\[
	\E\big[\langle U_N\rangle\big]= - \alpha_p(N)\,\E\big[g_{12\dots p}\langle S^z_{1}S^z_{2}\cdots S^z_{p} \rangle\big]\sqrt{\frac{N^{p+1}}{p!\,2}}
\] 
is now related to the mean of the $p$-th power of the replica-overlap operator
\[
	\E\big[\langle R_N^p \rangle^\otimes\big]=\alpha_p(N)\,\E\big[\langle S^z_{1}S^z_{2}\cdots S^z_{p}\rangle^2\big] + o_p(N)
\]
where $o_p(N)$ is a term which goes to zero as $N\to\infty$. Since the double commutator \eqref{eq:commutator} for $A=S^z_{1}S^z_{2}\cdots S^z_{p}$ equals $4\beta b\sum_{j=1}^p S_j^x$, we thus obtain the following generalization of Theorem~\ref{thm:QSK}:
\begin{theorem}
The mean of the $p$-th power of the replica-overlap operator is lower bounded according to
\begin{align}\label{eq:main2}
	\E\big[\langle R_N^p \rangle^\otimes\big] &\geq \alpha_p(N)\Phi\Big(p\beta b\,\E\big[\langle S_1^{x}\rangle\big]\Big)\notag\\
	&\quad + \frac{2}{\beta J}\frac{1}{N}\E\big[\langle U_N\rangle\big] + o_p(N)
\end{align}
for any $\beta>0$, $J>0$, $b\geq0$, and all $N\geq p$.
\end{theorem}
As before, we may further estimate the transverse magnetization, $\langle S_1^{x}\rangle\leq\tanh(\beta b)$, and bound the second term in \eqref{eq:main2} by the ground-state energy of the zero-field $p$-spin model, which itself is asymptotically (as $N\to\infty$) lower bounded by $-J\sqrt{\ln(2)}$, the known value for $p\to\infty$, using Slepian's lemma (see~\cite{B12}). This proves a spin-glass phase in a regime where the temperature and the field are low enough \cite{DT90,ONS07}. However, the larger we choose $p$, the smaller the regime becomes. In the limit $p\to\infty$ replica-symmetry breaking cannot be concluded by the above strategy.

This limit corresponds to the quantum random energy model (QREM). Its zero-field part $U_N$ is given in its (canonical) eigenbasis by the eigenvalues $- g_{\pmb{\sigma}}\sqrt{N/2}$ with standard Gaussian random variables $(g_{\pmb{\sigma}})$, which are independent and identically distributed for distinct $z$-configurations $\pmb{\sigma}\in\{-1,1\}^N$. In this case the phase diagram is known \cite{G90} for general $\beta$ and $b\geq0$, even at the rigorous level \cite{MW20}. As Goldschmidt's calculations \cite{G90} suggest, in the spin-glass phase the whole distribution of the replica overlap $\langle R_N\rangle^{\otimes}$ of the QREM turns out to agree with its classical analog. In particular, for this phase one can prove \cite{MW22} that $\lim_{N\to\infty}\E\big[\langle R_N\rangle^{\otimes}\big]=1-2\sqrt{\ln(2)}/(\beta J)$.\\

\paragraph{Conclusion.}

We have presented a simple argument that establishes replica-symmetry breaking in spin-glass models with a transverse field. It relies on a susceptibility bound from \cite{FB69} combined with an extension of the classical relation between the mean spin-glass order parameter $\overline{q}$ and the mean of the zero-field part of the energy to the quantum case. For the prominent quantum SK model, we have discussed in detail two resulting strictly positive but not optimal lower bounds on $\overline{q}$. Nevertheless, our method has extensions beyond the quantum SK model.

\begin{acknowledgments}
We thank one of the referees for stimulating us to consider also temperatures
between $J/(2\kappa)$ and $J$. 
CM and SW are supported by the DFG under EXC-2111--390814868. 
\end{acknowledgments}


\begin{thebibliography}{ONS07}

\bibitem{AB20}
A.~Adhikari and C.~Brennecke, \textit{Free energy of the quantum Sherrington--Kirkpatrick spin-glass model with transverse field}, J. Math. Phys. \textbf{61}, 083302, 16\,pp. (2020).

\bibitem{ALR87}
M.~Aizenman, J.~Lebowitz, and D.~Ruelle, \textit{Some rigorous results on the Sherrington--Kirkpatrick spin glass model}, Commun. Math. Phys. \textbf{112}, 3--20 (1987). Addendum \textbf{116}, 527 (1988).

\bibitem{AL18}
T.~Albash and D.\,A.~Lidar, \textit{Adiabatic quantum computation}, Rev. Mod. Phys. \textbf{90}, 015002, 64\,pp. (2018).

\bibitem{BL18}
C.\,L.~Baldwin and C.\,R.~Laumann, \textit{Quantum algorithm for energy matching in hard optimization problems}, 
Phys. Rev. B \textbf{97}, 224201, 19\,pp. (2018).

\bibitem{BL+17}
C.\,L.~Baldwin, C.\,R.~Laumann, A.~Pal, and A.~Scardicchio, \textit{Clustering of nonergodic eigenstates in quantum spin glasses}, Phys. Rev. Lett. \textbf{118}, 127201, 6\,pp. (2017).

\bibitem{B+13}
V.~Bapst, L.~Foini, F.~Krzakala, G.~Semerjian, and F.~Zamponi, \textit{The quantum adiabatic algorithm applied to random optimization problems: the quantum spin glass perspective}, Phys. Rep. \textbf{523}, 127--205 (2013).

\bibitem{BM80} 
A.\,J.~Bray and M.\,A.~Moore, \textit{Some observations on the mean-field theory of spin glasses}  J. Phys. C: Solid State Phys. \textbf{13} 419--434 (1980).


\bibitem{B12} A.~Bovier. \textit{Statistical Mechanics of Disordered Systems. A Mathematical Perspective} (Cambridge UP, Cambridge, 2006).


\bibitem{BU90} 
G.~B\"uttner and K.~D.~Usadel, \textit{Replica-symmetry breaking for the Ising spin glass in a transverse field}, Phys. Rev. B \textbf{42}, 6385--6395 (1990).

\bibitem{C+21}
A.~Callison, M.~Festenstein, J.~Chen, L.~Nita, V.~Kendon, and N.~Chancellor, \textit{Energetic perspective on rapid quenches in quantum annealing}, PRX Quantum \textbf{2}, 010338, 21\,pp. (2021).

\bibitem{C07}
N.~Crawford, \textit{Thermodynamics and universality for mean field quantum spin glasses}, Commun. Math. Phys. \textbf{274}, {821--839} (2007).

\bibitem{CR02}
A.~Crisanti and T.~Rizzo, \textit{Analysis of the $\infty$-replica symmetry breaking solution of the Sherrington--Kirkpatrick model}, Phys. Rev. E \textbf{65}, 046137, 9\,pp. (2002).

\bibitem{D85}
B.~Derrida, \textit{A generalization of the random energy model which includes correlations between energies}, J. Physique Lett. \textbf{46}, L401--L407 (1985).

\bibitem{DT90} 
V.~Dobrosavljevic and D.~Thirumalai, \textit{$1/p $ expansion for a $ p$-spin interaction spin-glass model in a transverse field}, J. Phys. A: Math. Gen. \textbf{23}, L767--L774 (1990).

\bibitem{DLS78}
F.\,J.~Dyson, E.\,H.~Lieb, and B.~Simon, \textit{Phase transitions in quantum spin systems with isotropic and nonisotropic interactions}, J. Stat. Phys. \textbf{18}, 335--383 (1978).

\bibitem{DA+15}
A.~Dutta, G.~Aeppli, B.\,K.~Chakrabarti, U.~Divakaran, T.\,F.~Rosenbaum, and D.~Sen, \textit{Quantum Phase Transitions in Transverse Field Spin Models -- From Statistical Physics to Quantum Information} (Cambridge University Press, Delhi, 2015).

\bibitem{FB69}
H.~Falk and L.\,W.~Bruch, \textit{Susceptibility and fluctuation}, Phys. Rev. \textbf{180}, 442--444 (1969).

\bibitem{FGGZ19}
E.~Farhi, J.~Goldstone, S.~Gutmann, and L.~Zhou, \textit{The quantum approximate optimization algorithm and the Sherrington--Kirkpatrick model at infinite size}, preprint arXiv:1910.08187 (2019).

\bibitem{FS86}
Ya.\,V.~Fedorov and E.\,F.~Shender, \textit{Quantum spin glasses in the Ising model with a transverse field}, JETP Lett. \textbf{43}, {681--684} (1986). [Russian original: Pis'ma Zh. Eksp. Teor. Fiz. \textbf{43}, {526--528} (1986)]

\bibitem{FH91}
K.\,H.~Fischer and J.\,A.~Hertz, \textit{Spin Glasses} (Cambridge University Press, Cambridge, 1991).

\bibitem{F65}
M.\,E.~Fisher, \textit{Bounds for the derivatives of the free energy and the pressure of a hard-core system near close packing}, J. Chem. Phys. \textbf{42} 3852--3856 (1965). 

\bibitem{G85}
E.~Gardner, \textit{Spin glasses with $ p $-spin interactions}, Nucl. Phys. B \textbf{257} {747--765} (1985).

\bibitem{G90}
Y.\,Y.~Goldschmidt, \textit{Solvable model of the quantum spin glass in a transverse field}, Phys. Rev. B \textbf{41}, {4858--4861} (1990).

\bibitem{GL90}
Y.\,Y.~Goldschmidt and P.-Y.~Lai, \textit{Ising spin glass in a transverse field: Replica-symmetry-breaking solution}, Phys. Rev. Lett. \textbf{64}, {2467--2470} (1990).

\bibitem{G01} F.~Guerra, \textit{Sum rules for the free energy in mean field spin glass models}, Fields Institute Communications \textbf{30} {161--170} (2001).


\bibitem{K16} S.~Knysh, \textit{Zero-temperature quantum annealing bottlenecks in the spin-glass phase}, Nat. Commun. \textbf{7}, 12370, 9\,pp. (2016).

\bibitem{KTH98}
R.~Kubo, M.~Toda, and N.~Hashitsume, \textit{Statistical Physics II -- Nonequilibrium Statistical Mechanics} (Springer, Berlin, 1998), 2nd edition, 3rd corrected printing.

\bibitem{LPS14}
C.\,R.~Laumann, A.~Pal, and A.~Scardicchio, \textit{Many-body mobility edge in a mean-field quantum spin glass}, Phys. Rev. Lett. \textbf{113}, 200405, 5\,pp. (2014).

\bibitem{LRRS21}
H.~Leschke, S.~Rothlauf, R.~Ruder, and W.~Spitzer, \textit{The free energy of a quantum Sherrington--Kirkpatrick spin-glass model for weak disorder}, J. Stat. Phys. \textbf{182}, 55, 41\,pp. (2021). 

\bibitem{MW20}
C.~Manai and S.~Warzel, \textit{Phase diagram of the quantum random energy model}, J. Stat. Phys. \textbf{180}, {654--664} (2020).

\bibitem{MW21}
C.~Manai and S.~Warzel, \textit{Generalized random energy models in a transversal magnetic field: free energy and phase diagrams}, preprint arXiv:2007.03290, 30\,pp. (2020)\,(to appear in Probab. Math. Phys.).

\bibitem{MW22}
C.~Manai and S.~Warzel (in preparation)

\bibitem{MM09}
M.~M\'ezard and A.~Montanari, \textit{Information, Physics, and Computation} (Oxford UP, Oxford, 2009).

\bibitem{M21}
A.~Montanari, \textit{Optimization of the Sherrington--Kirkpatrick Hamiltonian}, SIAM J. Comput. \textbf{0} (0), FOCS19-1--FOCS19-38 (2021).

\bibitem{MRC18}
S.~Mukherjee, A.~Rajak, and B.\,K.~Chakrabarti, \textit{Possible ergodic-nonergodic regions in the quantum Sherrington--Kirkpatrick spin glass model and quantum annealing}, Phys. Rev. E \textbf{97}, 022146, 6\,pp. (2018).

\bibitem{MC19}
S.~Mukherjee and B.\,K.~Chakrabarti, \textit{On the question of ergodicity in quantum spin glass phase and its role in quantum annealing},  J. Phys. Soc. Jpn. \textbf{88}, 061004, 10\,pp. (2019).

\bibitem{N01} 
H.~Nishimori, \textit{Statistical Physics of Spin Glasses and Information Processing -- An Introduction} (Clarendon, Oxford, 2001).

\bibitem{ONS07} 
T.~Obuchi, H.~Nishimori, and D.~Sherrington, \textit{Phase diagram of the p-spin-interacting spin glass with ferromagnetic bias and a transverse field in the infinite-$p$ limit}, J. Phys. Soc. Jpn. \textbf{76},  054002, 10\,pp. (2007).

\bibitem{P08}
D.~Panchenko, \textit{On differentiability of the Parisi formula}, Elect. Comm. in Probab. \textbf{13}, {241--247} (2008).

\bibitem{P80a}
G.~Parisi, \textit{The order parameter for spin glasses: a function on the interval 0--1}, J. Phys. A: Math. Gen. \textbf{13}, {1101--1112} (1980).

\bibitem{P80b}
G.~Parisi, \textit{A sequence of approximated solutions to the S--K model for spin glasses}, J. Phys. A: Math. Gen. \textbf{13}, {L115--L121} (1980).

\bibitem{RCC89}
P.~Ray, B.\,K.~Chakrabarti, and A.~Chakrabarti, \textit{Sherrington--Kirkpatrick model in a transverse field: Absence of replica symmetry breaking due to quantum fluctuations}, Phys. Rev. B \textbf{39}, 11828--11832 (1989).

\bibitem{R77}
G.~Roepstorff, \textit{A stronger version of Bogoliubov's inequality and the Heisenberg model}, Commun. Math. Phys. \textbf{53}, 143--150 (1977).

\bibitem{SK75} D.~Sherrington and S.~Kirkpatrick, \textit{Solvable model of a spin-glass}, Phys. Rev. Lett. \textbf{35}, 1792--1796 (1975).

\bibitem{S+20}
V.\,N.~Smelyanskiy, K.~Kechedzhi, S.~Boixo, S.\,V.~Isakov, H.~Neven, and B.~Altshuler, \textit{Nonergodic delocalized states for efficient population transfer within a narrow band of the energy landscape}. Phys. Rev X \textbf{10}, 011017, 51\,pp. (2020). 

\bibitem{S85} 
H.-J. Sommers, \textit{Parisi function $ q(x) $  near $ T_{\textnormal{c}}$}, J. Physique Lett. \textbf{46}, L779--L785 (1985). 

\bibitem{S+13}
S.~Suzuki, J.-i.~Inoue, and B.\,K.~Chakrabarti, \textit{Quantum Ising Phases and Transitions in Transverse Ising Models} (Springer, Berlin, 2013), 2nd ed.


\bibitem{T06} M.~Talagrand, \textit{The Parisi formula}, Ann. Math. \textbf{163}, {221--263} (2006).

\bibitem{T06c} M.~Talagrand, \textit{Parisi measures}, J. Funct. Anal. \textbf{231}, {269--286} (2006).



\bibitem{US87} K.\,D.~Usadel and B.~Schmitz, \textit{Quantum fluctuations in an Ising spin glass with transverse field}, Solid State Commun. \textbf{64}, {975--977} (1987).

\bibitem{YI87}
T.~Yamamoto and H.~Ishii, \textit{A perturbation expansion for the Sherrington--Kirkpatrick model with a transverse field}, J. Phys. C \textbf{20}, 6053--6060 (1987).

\bibitem{Y17} 
A.\,P.~Young, \textit{Stability of the quantum Sherrington--Kirkpatrick spin glass model}, Phys. Rev. E \textbf{96}, 032112, 6\,pp. (2017).

\end{thebibliography}
\end{document}